\documentclass[12pt,preprint]{aastex}

\newcommand{\teff}{$T_{\mathrm{eff}} \;$}
\newcommand{\tef}{$T_{\mathrm{eff}}$}
\newcommand{\feh}{[\mathrm{Fe/H}]}
\newcommand{\trip}{\ion{O}{1} triplet }

\begin{document}

\title{Oxygen from the $\lambda7774$ High-Excitation Triplet in Open Cluster 
Dwarfs: Hyades\altaffilmark{1,2}}

\altaffiltext{1}{Based on observations obtained with the Mayall 4-m telescope at 
Kitt Peak National Observatory, a division of the National Optical Astronomy
Observatories, which is operated by the Association of Universities for Research
Astronomy, Inc. under cooperative agreement with the National Science
Foundation.}
\altaffiltext{2}{This paper includes data taken with the Harlan J. Smith 2.7-m 
telescope at The McDonald Observatory of the University of Texas at Austin.}

\author{Simon C. Schuler AND Jeremy R. King}
\affil{Department of Physics and Astronomy, Clemson University}
\affil{118 Kinard Laboratory, Clemson, SC, 29634}
\email{sschule,jking2@ces.clemson.edu}
\author{Donald M. Terndrup AND Marc H. Pinsonneault}
\affil{Department of Astronomy, The Ohio State University}
\affil{140 West 18th Avenue, Columbus, OH, 43210}
\email{terndrup,pinsono@astronomy.ohio-state.edu}
\author{Norman Murray}
\affil{Canadian Institute for Theoretical Astrophysics, University of Toronto}
\affil{60 St. George Street, Toronto, ON, M5S3H8, Canada}
\email{murray@cita.utoronto.ca}
\and
\author{L.M. Hobbs}
\affil{University of Chicago}
\affil{Yerkes Observatory, Williams Bay, WI, 53191}
\email{hobbs@yerkes.uchicago.edu}

\begin{abstract}
Oxygen abundances have been derived from the near-IR, high-excitation 
$\lambda 7774$ \ion{O}{1} triplet in high-resolution, high signal-to-noise 
spectra of 45 Hyades dwarfs using standard one dimensional, plane-parallel LTE 
models.  Effective temperatures of the stellar sample range from
$4319-6301$ K, and the derived relative O abundances as a function of \teff 
evince a trichotomous morphology.  At $T_{\mathrm{eff}} > 6100 \; 
\mathrm{K}$, there is evidence of an increase in the O abundances with 
increasing \tef, consistent with non-LTE (NLTE) predictions.  At intermediate 
\teff ($5450 \leq T_{\mathrm{eff}} \leq 6100 \; \mathrm{K}$), the O abundances 
are flat, and star-to-star values are in good agreement, having a mean value of 
$\mathrm{[O/H]} = +0.25 \pm 0.02$; however, systematic errors at the $\lesssim 
0.10 \; \mathrm{dex}$ level might exist.  The O abundances for stars with 
$T_{\mathrm{eff}} \leq 5450 \; \mathrm{K}$ show a striking increase with 
{\it decreasing} \tef, in stark contrast to expectations and canonical NLTE 
calculations.  The cool Hyades triplet results are compared to those recently 
reported for dwarfs in the Pleiades cluster and the UMa moving group; qualitative 
differences between the trends observed in these stellar aggregates point to a 
possible age-related diminution of triplet abundance trends in cool open cluster 
dwarfs.  Correlations with age-related phenomena, i.e., chromospheric activity 
and photospheric spots, faculae, and/or plages, are investigated.  No correlation 
with \ion{Ca}{2} H+K chromospheric activity indicators is observed.  
Multi-component LTE ``toy'' models have been constructed in order to simulate 
photospheric temperature inhomogeneities that could arise from the presence of 
starspots, and we demonstrate that photospheric spots are a plausible source of 
the triplet trends among the cool dwarfs.  
\end{abstract}


\section{INTRODUCTION}
The derivation of stellar O abundances using spectra at visible wavelengths 
is generally limited to a small number of atomic features.  The two sets of
features most often used for O abundance determinations are a) two forbidden 
[\ion{O}{1}] spectral lines at 6300.3 and 6363.8 {\AA}, and b) the 
permitted high-excitation \ion{O}{1} triplet in the 7771-7775 {\AA} region.  The 
formation of the [\ion{O}{1}] lines is well described by current local 
thermodynamic equilibrium (LTE) calculations \citep{2003A&A...402..343T}, but 
the lines are weak in the spectra of solar-type stars and suffer from blends
(e.g., Johansson et al. 2003).  High-quality spectra and knowledge of the 
blending features are required if the forbidden lines are to be used for 
accurate abundance derivations.   On the other hand, each line of the \trip is 
strong in the spectra of solar-type dwarfs and evolved stars at all 
metallicities, and they are presumably free from blends, making them a more 
attractive choice for O abundance studies.  However, the \trip (henceforth the 
triplet) is known to be sensitive to non-LTE (NLTE) effects 
\citep{1991A&A...245L...9K}, and there is much controversy regarding its delivery
of accurate O abundances. 

The formation of the triplet is due to a high-excitation electronic transition
(9.15 eV above the ground state) from the 3s orbital into the 3p orbital, with 
the multiple lines of the feature resulting from three possible values of the 
total angular momentum ($J$) in the terminal state.  The NLTE behavior of the 
triplet is due to the dilution of each line's source function ($S_l$) compared to
the Planck function in the line forming region \citep{2001NewAR..45..559K}.  
Thus, O abundances derived using LTE calculations, which utilize a Planckian 
source function, are overestimated.  The magnitude of the effect is expected to 
increase as the number of O atoms in the 3s state increases, for instance by an 
increase in the total number of O atoms, by a decrease in gas pressure, or by an 
increase in temperature in the line forming region (e.g., Takeda 2003, Kiselman 
2001).  In particular, \citet{2003A&A...402..343T} has constructed an extensive
collection of NLTE corrections for late-F through early-K stars in various
evolutionary states with the general result that the NLTE corrections for a 
given adopted O abundance increase with increasing \teff for all evolutionary 
states, with negligible corrections required for dwarfs with $T_{\mathrm{eff}} < 
6000 \; \mathrm{K}$.

On the observational side, quantifying the NLTE behavior of derived triplet
abundances is difficult because of the inhomogeneity of O abundances among field
stars; star-to-star comparisons are generally not meaningful.  This can be
overcome by deriving simultaneously O abundances from the triplet and a feature
predicted to be free from NLTE effects, such as $\lambda6300$ [\ion{O}{1}] line. 
For example, \citet{1995AJ....109..383K} compared LTE O abundances as derived 
from the triplet and the $\lambda 6300$ [\ion{O}{1}] in a sample of field dwarfs 
with a moderate range in metallicities ($-0.84 \leq \mathrm{[Fe/H]} \leq +0.24$; 
[Fe/H] is used throughout this paper as a fiducial for metallicity).  No 
significant difference in the O abundance from the two indicators was found for 
stars with $T_{\mathrm{eff}} \lesssim 6200 \; \mathrm{K}$, but an increasing 
difference- presumably due to larger triplet abundances- with increasing \teff 
was evident in the warmer stars of the sample.  The results of the King \& 
Boesgaard study are in good agreement with NLTE predictions.  Many other studies 
have derived NLTE O abundances using the triplet, with varying results.  
Additional observational behavior of the triplet compared to other O spectral 
lines has been covered extensively in the literature (e.g., Fulbright \& Johnson 
2003; Nissen et al. 2002; Mishenina et al. 2000; Israelian, Garc{\'i}a 
L{\'o}pez, \& Rebolo 1998), and we point the reader interested in a more 
thorough discussion to those publications.  

Another tactic to investigate triplet departures from LTE is to analyze O
abundances of open cluster stars.  Open clusters are stellar aggregates composed 
of members believed to have a common age and chemical composition.  These shared 
characteristics make {\it intracluster} comparisons ideal for probing 
mass-dependent stellar effects and {\it intercluster} comparisons ideal for 
probing age-related or metallicity effects-. Surprisingly, only 
a handful of cluster O studies have been done.  \citet{1993ApJ...412..173G} 
subjected the triplet in spectra of 25 Hyades (600 Myr, $\mathrm{[Fe/H]} = 
+0.13$) F dwarfs ($6045 \leq T_{\mathrm{eff}} \leq 7375 \; \mathrm{K}$), as well 
as 18 F dwarfs ($5850 \leq T_{\mathrm{eff}} \leq 7110 \; \mathrm{K}$) in the Ursa 
Major (UMa) kinematic moving group (600 Myr, $\mathrm{[Fe/H]} = -0.09$), to an 
NLTE analysis.  While good star-to-star agreement (within uncertainties) of the O 
abundances for both stellar samples was found, \tef-dependent differences are 
seen when the stars are placed into \teff bins.  The mean bin O abundance 
decreases with decreasing \teff for both the Hyades and UMa, raising questions as 
to the  accuracy of the adopted NLTE corrections.  \citet{kphd} analyzed the 
triplet in the spectra of four Hyades dwarfs ($5772 \leq T_{\mathrm{eff}} \leq 
6103 \; \mathrm{K}$) and found star-to-star agreement in the LTE O abundances.  
\citet{2000ApJ...533..944K} derived LTE O abundances from the triplet in two 
Pleiades (100 Myr, $\mathrm{[Fe/H]} = 0.00$) and one NGC 2264 (10 Myr, 
$\mathrm{[Fe/H]} = -0.15$) K-dwarfs.  Given the relatively low \teff for these 
stars ($4410 \leq T_{\mathrm{eff}} \leq 4660 \; \mathrm{K}$), NLTE effects on the 
formation of the triplet are predicted to be negligible 
\citep{2003A&A...402..343T}, yet unexpectedly high O abundances were found for 
all three stars.  No explanation for the high abundances was immediately obvious, 
but the authors speculated that yet unknown NLTE effects resulting from the 
presence of a chromosphere may be the cause (as suggested by Takeda 1995).  

A more comprehensive study of LTE O abundances derived using the triplet in 
Pleiades dwarfs, as well as in M34 (200 Myr, $\mathrm{[Fe/H]} = +0.07$) dwarfs, 
has been presented by Schuler et al. (2004; henceforth SKHP).  The stellar sample 
includes 15 Pleiads with $5048 \leq T_{\mathrm{eff}} \leq 6172 \; \mathrm{K}$ and eight M34 
dwarfs with $5385 \leq T_{\mathrm{eff}} \leq 6130 \; \mathrm{K}$, and in both cases, a 
dramatic increase in triplet [O/H] abundances with {\it decreasing} \teff is 
seen for $T_{\mathrm{eff}} \lesssim 5600 \; \mathrm{K}$.  The Pleiades results 
confirmed the high triplet-based O abundances initially reported by 
\citet{2000ApJ...533..944K} and are in contrast with canonical NLTE predictions. 
Furthermore, the Pleiades abundances evince scatter that exceeds uncertainties 
at $T_{\mathrm{eff}} < 5300 \; \mathrm{K}$.  Following 
\citet{1995PASJ...47..463T}, the triplet results were compared to $H\alpha$ and 
\ion{Ca}{2} triplet chromospheric activity indicators from the literature.  No 
correlations between O abundances and activity or O abundance scatter and 
activity scatter were found; it should be noted, however, that the O abundances
and activity indices were not measured contemporaneously.  SKHP proposed 
temperature inhomogeneities resulting from photospheric spots, faculae, and/or 
plages (henceforth collectively referred to as spots) as a possible explanation 
for the high O abundances, being influenced by the work of 
\citet{2003AJ....126..833S} who suggested the anomalously blue color excess of 
Pleiades K-dwarfs may be due to the presence of spots.  \citet{ks05} extended 
the study of the triplet to six UMa dwarfs in the \teff range of 4925-5827 K, 
and the LTE O abundances show a similar increase with decreasing \teff as seen 
in the Pleiades.  The UMa trend, however, is shallower.  The age and metallicity 
differences between the Pleiades and UMa suggest that triplet abundance trends
observed among cool open cluster dwarfs might be influenced by effects related 
to either of these parameters.

The Hyades open cluster is the next logical target for our continued
investigation of the ubiquity and nature of triplet-derived O abundance trends 
in cool cluster dwarfs.  The cluster is observationally appealing because of the
relative brightness of its members; late-K dwarfs are brighter than $V \sim 10.5$
and thus within the capabilities of moderate-sized telescopes.  More importantly,
the physical characteristics are well-suited for this study.  Its members are 
metal-rich compared to those in both the Pleiades and UMa, and they have an age 
that is approximately coeval to UMa members \citep{ks05}.  If a Hyades triplet 
trend is present, comparing its morphology to that of the Pleiades and UMa 
should provide valuable insight into its possible cause and evolution.  
Therefore, an LTE analysis of the triplet in high-resolution, high 
signal-to-noise (S/N) spectra obtained with the Mayall 4-m telescope and echelle 
spectrograph at Kitt Peak National Observatory and with the Harlan J. Smith 
2.7-m telescope and the cross-dispersed echelle spectrometer at McDonald 
Observatory of 45 Hyades dwarfs is presented.  The cool dwarf O abundances 
derived from the triplet are compared to previous open cluster results, and a 
plausible explanation for observed triplet trends is proffered.

\section{OBSERVATIONS AND DATA REDUCTION}
Hyades membership is well determined (Perryman et al. 1998), and the
cluster is rich with G and K dwarfs.  We made use of the recent studies of 
\citet{2004ApJ...603..697Y} and \citet{paulson} in constructing our 
target list, which contains 45 stars.  The analysis presented herein makes use 
of three sets of high-resolution echelle spectra obtained with two telescopes on 
different dates.  The majority of the sample were observed 2002 November 22-25 
with the KPNO Mayall 4-m and accompanying Cassegrain echelle spectrograph.  This 
sample will henceforth be referred to as KPNO-02.  The $31.6-63 \; \mathrm{g} \; 
\mathrm{mm}^{-1}$ echelle grating and $226-1 \; \mathrm{g} \; \mathrm{mm}^{-1}$ 
cross disperser were used, resulting in a dispersion of $\sim 0.091 \; 
\mathrm{\AA} \; \mathrm{pixel}^{-1}$ and a resolution $R = \lambda / \Delta 
\lambda \approx 43,000$ ($\sim 2.0 \; \mathrm{pixels}$) at 7775 {\AA}.  The 
detector (T2KB) consists of $24 \; \mu \mathrm{m}$ pixels and has dimensions of 
$2048 \times 2048$; no binning was used.  

The second set of spectra was also obtained with the KPNO Mayall 4-m telescope 
and echelle spectrograph; the observations took place on the dates of 2004 
December 20-23 (henceforth KPNO-04).  These later observations utilized the 
$58.5-63 \; \mathrm{g} \; \mathrm{mm}^{-1}$ echelle grating and the $226-1 \; 
\mathrm{g} \; \mathrm{mm}^{-1}$ cross disperser.  The data are characterized by 
a dispersion of $\sim 0.093 \; \mathrm{\AA} \; \mathrm{pixel}^{-1}$  and a 
similar resolution of $R \approx 42,000$ ($\sim 2.0 \; \mathrm{pixels}$) at 7775 
{\AA}.  The T2KB detector was again used without binning.

The final set of spectra was obtained on 2004 October 10-12 with the
Harlan J. Smith 2.7-m telescope and the "2dcoude" cross-dispersed echelle 
spectrometer at McDonald Observatory (henceforth McD-04).  The cs23 setting and 
E2 echelle grating ($52.7 \mathrm{g} \; \mathrm{mm}^{-1}$) were used, giving a 
resolving power of $R\approx 60,000$ ($\sim 2.1 \; \mathrm{pixels}$) and 
dispersion of $\sim 0.063 \; \mathrm{\AA} \; \mathrm{pixel}^{-1}$ at 7775 {\AA}. 
The angle of the echelle grating and cross dispersion prism was set 
manually so that the \ion{O}{1} triplet fell well away from the edge of the 
detector.  The TK3 detector, with $24 \; \mu \mathrm{m}$ pixels, is a $2048 
\times 2048$ square ccd, and it was used without binning.  We note that one star 
(HIP 20146) was kindly acquired by Dr. J. Laird with the McDonald 2.7-m 
telescope in 1995 with a similar set-up as the 2004 2.7-m observations and is 
included in our sample.  Its spectrum is of equal quality to the other McDonald 
data.

The stars in the combined sample are listed according to Hipparcos number, 
along with the HD number when available, in Table 1.  The adopted 
stellar parameters (see below), as well as the telescope used to observe each 
object, are also presented therein.  It will be noticed that a handful of stars 
have been observed during two different runs.  This is beneficial in that it 
allows for the revelation of possible systematic errors that might be associated 
with a particular data set.  A solar proxy spectrum of the daytime sky was 
obtained during each observing run, as is indicated in Table 1.  The KPNO and 
McDonald data were reduced in a consistent fashion using 
{\sf IRAF}\footnotemark[3], following standard procedures of bias subtraction, 
flat fielding, correction for scattered light, extraction, and wavelength 
calibration.  A sample spectrum from each data set is shown in Figure 1.
\marginpar{Tab.~1}
\marginpar{Fig.~1}

\footnotetext[3]{IRAF is distributed by the National Optical Astronomy 
Observatories, which are operated by the Association of Universities for 
Research in Astronomy, Inc., under cooperative agreement with the National 
Science Foundation.}

\section{ANALYSIS AND RESULTS}
\subsection{Stellar Parameters}
Effective temperatures have been derived photometrically following the procedure 
of \citet{1993ApJ...415..150T}.  Briefly, ($B-V)_0$ colors are transformed to 
($V-K)_0$ colors using an empirical relation for Hyades dwarfs \citep{carney}.  
Effective temperatures are then derived using the ($V-K$)-\teff relation of 
\citet{carney} modified to include the zero-point correction of 
\citet{1985A&A...146..249C}.  The $1\sigma$ error in the resulting temperatures 
as reported by \citet{1993ApJ...415..150T} is $\pm 55 \; \mathrm{K}$.   The 
($B-V$) colors for the largest number of stars are from 
\citet{1999A&A...352..555A}, and for the stars not appearing therein, the colors
are from \citet{paulson} or \citet{2004ApJ...603..697Y}.

Surface gravities ($\log g$) have been determined by making use of the latest
$\mathrm{Y}^2$ isochrones \citep{y2} and the interpolation routine
provided by the $\mathrm{Y}^2$ consortium\footnotemark[4] to generate a stellar 
track suitable for the Hyades.  The required cluster parameters for the 
interpolation are age, [Fe/H], and $\alpha$-enrichment ([$\alpha$/Fe]).  There 
have been multiple studies on the age and Fe content of the Hyades.  We have 
chosen an age of 600 Myr (e.g., Perryman et al. 1998; Torres, Stefanik, \& Latham 
1997) and a relative Fe abundance of $\feh =+0.13$ \citep{paulson}.  Paulson et 
al. also derived the abundances of the $\alpha \; \mathrm{elements}$ Mg, Si, Ca, 
and Ti and found them to scale with Fe; thus the Hyades $\alpha$-enrichment has 
been taken to be zero.  Surface gravities and effective temperatures taken from 
the stellar track characterized by the above parameters were fit with a fourth 
order Legendre polynomial using the {\sf curfit} utility within {\sf IRAF}.  
This utility provides equivalent power series coefficients which were used to 
create a $T_{\mathrm{eff}}$-dependent relation for $\log g$.  The final $\log g$ 
values were calculated using the adopted $T_{\mathrm{eff}}$ described above. 
Additionally, the effect of the adopted stellar track parameters on the derived 
surface gravities has been tested by creating several additional stellar tracks 
interpolated with a mixture of relative iron abundances and ages ranging from 
$0.10 \leq \feh \leq 0.16$ and 600-700 Myr.  The surface gravities derived from 
these different tracks never differ from the adopted values by more the 0.01 
dex, demonstrating the insensitivity of the $\log g$ values for these
main sequence (MS) dwarfs to the reasonable choice of cluster parameters.  We 
have chosen a conservative $1\sigma$ error in $\log g$ of 0.10 dex.

\footnotetext[4]{See http://www.astro.yale.edu/demarque/yyiso.html}

Microturbulent velocities ($\xi$) are calculated using the empirical formula of 
\citet{2004A&A...420..183A}.  The function predicts $\xi$ values with an rms 
scatter of $0.14 \; \mathrm{km \; s}^{-1}$.  We have chosen a $1\sigma$ uncertainty 
of $0.15 \; \mathrm{km \; s}^{-1}$ in our $\xi$ values.   The adopted stellar 
parameters are given in Table 1.

\subsection{Equivalent Widths}
Continuum normalization and equivalent width (EW) measurements were carried out
using the one-dimensional spectrum analysis package {\sf SPECTRE} (Fitzpatrick \&
Sneden 1987).  The triplet falls in two separate orders of the KPNO-02 
data due to order overlap in the individual spectra.  EWs of the triplet have 
been measured in both orders- designated as blue and red- and abundances are 
derived using the blue, red, and mean EWs.  Gaussian profiles were used to
determine the EWs of the triplet feature; the three lines were measurable in the 
spectra of all the stars except for eight:  two lines were measurable for seven 
stars and one line for one star.  These eight stars are generally the coolest 
stars in the sample as the triplet feature becomes increasingly weak in the 
spectra of dwarfs with decreasing temperature, ceteris paribus.  The dependence 
of triplet EWs on \teff is clearly seen in Figure 2.  Only stars for which all 
three lines were measurable are plotted in the figure, and thus all the stars 
cooler than $\sim 4800 \; \mathrm{K}$ are not included.  Also apparent in Figure 
2 is the uniformity of EW measurements at a given \teff shared among the 
different data sets.  This is encouraging and suggests our individual samples do 
not suffer from observation-dependent systematic errors.  The measured EWs and 
the per pixel signal-to-noise (S/N) ratio, as given by Poisson statistics, in 
the $\lambda 7775$ region are given in Tables 2 and 3.  Some of the stars were 
observed multiple times during a given run, and the EW and S/N measurements for 
these were made using the co-added spectra.
\marginpar{Fig.~2}
\marginpar{Tab.~2}

Empirical estimation of line strength uncertainties has been made by comparing 
the EWs from the blue and red orders of the KPNO-02 data.  Line-by-line 
differences were calculated for each star in the KPNO-02 sample, and the mean 
difference for each line has been adopted as the $1\sigma$ uncertainty in the 
EW of the line.  The final uncertainties are 4.4 m{\AA} for $\lambda 7771.94$, 
4.8 m{\AA} for $\lambda 7774.17$, and 2.7 m{\AA} for $\lambda 7775.39$ and have 
been adopted for the entire stellar sample.  It should be noted that 35 out of 
36 stars comprising the KPNO-02 data set have effective temperatures greater 
than 4800 K and thus have larger EWs than the cooler stars in the sample.  The
adopted line strength uncertainties represent a far larger percentage of the EWs
of the weaker features in the coolest stars and may not accurately reflect the 
errors associated with their measurement.  Indeed, comparing the line 
strengths from the blue and red orders of the coolest star in the KPNO-02 sample
(HIP 18322) gives uncertainties of 0.4 and 0.7 m{\AA} for the $\lambda 7774.17$
and $\lambda 7775.39$ lines, respectively; the $\lambda 7771.94$ line was not
measurable for this star.  Nonetheless, we conservatively adopt the larger EW
uncertainties for the whole stellar sample.  The final uncertainties in the
derived abundances for the coolest stars are clearly dominated by the line
strength errors and are large relative to the warmer stars; however, the final
conclusions of this work remain unchanged.

\subsection{Oxygen Abundances}
Oxygen abundances have been derived following the procedure of SKHP using an 
updated version of the LTE stellar line analysis software package {\sf MOOG} 
(Sneden 2004, private communication).  Atomic data ($\chi_{7772} = \chi_{7774} = 
\chi_{7775} = 9.15 \; \mathrm{eV}; \log{gf}_{7772} = 0.369, \log{gf}_{7774} = 
0.223, \mathrm{and} \log{gf}_{7775} = 0.001$) are from the 
NIST\footnotemark[5] database.  Stellar atmosphere models characterized by the 
adopted stellar parameters and $\mathrm{[Fe/H]} = +0.13$ \citep{paulson} were 
interpolated from the ATLAS9 (LTE) grids of R. Kurucz\footnotemark[6].  The 
grids utilized here include the convective overshoot approximation.  The 
validity of using the overshoot models of Kurucz to derive Fe abundances for 
Hyades dwarfs has recently been questioned \citep{paulson}, who suggest Kurucz 
grids without convective overshoot are more appropriate.  However, SKHP derived 
triplet abundances in Pleiades and M34 dwarfs utilizing four different sets of 
ATLAS9 grids, including Kurucz grids with and without overshoot, and found the 
results to be independent of model atmosphere.  Models with convective overshoot 
have been used here due to the availability of finer super-solar metallicity 
steps for these grids compared to grids without the overshoot approximation.

\footnotetext[5]{http://physics.nist.gov/PhysRefData/ASD/index.html}
\footnotetext[6]{See http://kurucz.harvard.edu/grids.html}

The derived mean O abundances and final internal uncertainties are presented in 
Table 4; the abundances are given relative to solar values (using the usual notation 
$\mathrm{[O/H]} = \log N(\mathrm{O})_{Star} - \log N(\mathrm{O})_{\sun}$ 
on  a scale where $\log N(\mathrm{H}) = 12.0$) via a line-by-line 
comparison.  Solar abundances were derived in the same manner as the rest of the 
sample from spectra obtained during each observing run; each solar spectrum 
was used for comparison with only those stars observed during the same
observing run.  Abundance sensitivities to the adopted stellar parameters were 
determined by constructing additional model atmospheres characterized by single 
parameter changes of $\pm 150 \; \mathrm{K}$ in $T_{\mathrm{eff}}$, $\pm 0.25 \; 
\mathrm{dex}$ in $\log g$, and $\pm 0.30 \; \mathrm{km} \; \mathrm{s}^{-1}$ in 
$\xi$ for each star and deriving the adjusted abundances in the same manner as 
above; typical sensitivities are given in Table 5.  The resulting 
parameter-dependent abundance uncertainties are summed quadratically with 
uncertainties in line strengths and in the mean abundances to achieve the final 
internal uncertainties.
\marginpar{Tab.~3}
\marginpar{Tab.~4}

In Figure 3 the relative O abundances and associated error bars, as well as the
line-by-line relative abundances for each star, are plotted against 
$T_{\mathrm{eff}}$.  The KPNO-02 abundances are those derived using the mean EWs. 
The error bars for the coolest stars are significantly larger than those for the 
warmer stars due to the large adopted EW uncertainties relative to the cool object 
line strengths, as described above.  Regardless, a smooth increase in [O/H] vs 
\teff is clearly seen for stars with $T_{\mathrm{eff}} < 5500 \; \mathrm{K}$, as 
well as an apparent increase for stars with $T_{\mathrm{eff}} > 6200 \; 
\mathrm{K}$.  Also apparent is the lack of star-to-star scatter in the [O/H] 
abundances, in contrast to that observed among cool Pleiades dwarfs (SKHP).  
Abundances for stars that are common to more than one data set are given in 
Table 6.  Comparing results from the different data sets reveals no statistically 
significant deviations.  This was suggested by the uniformity of EWs in Figure 2, 
and more strongly demonstrated here.  We are confident that no large systematic 
differences exist between the data sets, and they will no longer be 
distinguished.  
\marginpar{Fig.~3}
\marginpar{Tab.~5}

There are also no systematic differences in the line-by-line O abundances (Figure
3), which are in good agreement for each star and provide strong evidence that
the triplet is not hampered by blends.  Possible exceptions are the two 
coolest stars, HIP 19441 and HIP 20762.  Only two of the three lines of the 
triplet were measurable for these two stars, and in each case, the abundance
of the $\lambda 7774$ line is $\sim 0.20$ dex larger than that of the other 
measurable line.  There is an \ion{Fe}{1} line at $\lambda = 7774.00 \; 
\mathrm{\AA}$ ($\chi = 5.01 \; \mathrm{eV}$) \citep{1998PASJ...50...97T} that may 
be enhancing the line strength of the $\lambda 7774$ O feature in the spectra of
the coolest stars, and we have used {\sf MOOG} to synthesize the triplet region 
of HIP 19441 in order to determine the magnitude of this effect.  According to the relative line strength 
parameter provided by {\sf MOOG}, the O line is expected to be an order of 
magnitude stronger than the Fe line, suggesting that this particular Fe feature 
is not responsible for the enhanced EW of the $\lambda 7774$ O line.  
Alternatively, systematic errors related to measuring the weak triplet in the 
spectra of the coolest dwarfs or a yet to be identified blend may be the cause of 
the seemingly higher abundance provided by the $\lambda 7774$ line.  Regardless, 
if the abundance of the $\lambda 7774$ feature is ignored for the two coolest 
stars, the morphology of the cool dwarf abundance trend is not greatly affected.

One star, HIP 15310, has an O abundance that deviates from those of stars in the
sample with similar \teff ($\sim 5870 \; \mathrm{K}$).  \citet{paulson} reports a 
Fe abundance for HIP 15310 that is larger than the cluster mean and introduced 
the possibility that this star might have been chemically enriched, possibly by 
the accretion of proto-planetary or planetary material.  The observational 
evidence that host stars of the discovered planetary systems are statistically 
more Fe-rich compared to otherwise similar field stars that do not have planets 
is now well established \citep{2005ApJ...622.1102F}.  Two theories to explain the 
higher metallicity of planetary hosts have been discussed in the literature: a) 
planets form preferentially in high-metallicity environments, or b) planetary 
hosts are chemically enriched due to planetary accretion.  
\citet{1997MNRAS.285..403G} suggested that a possible signature of accretion 
would be abundance enhancements correlated with condensation temperature.  
Oxygen is a volatile element with a condensation temperature of $\sim 180 \; 
\mathrm{K}$ (in a gas of solar composition), and Fe is a refractory element with 
a condensation temperature of $\sim 1340 \; \mathrm{K}$ 
\citep{2003ApJ...591.1220L}.  While a larger fraction of the initial O abundance 
of a gas condenses into rocky material than of other volatile elements- due to 
its role in the formation of silicates and oxides- the increase in its abundance 
in a stellar photosphere compared to refractory elements such as Fe should not 
be as significant if accretion has taken place.  Adopting a mean Hyades O 
abundance using the stars in the \teff range of 5450-6100 K (the \teff range 
over which there is good agreement in O abundances; see below) and the mean 
cluster Fe abundance from \citet{paulson}, a cluster [O/Fe] ratio of +0.12 is 
attained.  Paulson et al. derived an Fe abundance of $\mathrm{[Fe/H]}=+0.30$ for 
HIP 15310, and comparing this to the O abundance found here, $\mathrm{[O/Fe]} = 
+0.08$ for this star.  From these results, it is apparent that HIP 15310 has not 
been chemically enriched by fractionated material, although more quantitative 
data on the expected relative O and Fe enhancements due to accretion are in 
order.  Based on previous membership studies, \citet{paulson} also question the 
cluster membership of HIP1 15310.  Further investigation of HIP 15310 and its 
possible chemical enrichment is beyond the scope of this paper but certainly 
should be addressed in future studies.

\section{DISCUSSION}
Although the [O/H] abundances of the warmest stars ($T_{\mathrm{eff}} > 6150 \;
\mathrm{K}$) in the sample agree with those of slightly cooler stars within the 
calculated uncertainties (Figure 3), the abundances appear to be increasing with 
increasing \tef.  This behavior is consistent with previous triplet LTE abundance
derivations of near-solar metallicity dwarfs \citep{1995AJ....109..383K} and 
with expectations from canonical NLTE calculations (e.g., Takeda 2003).  More 
interesting is the trend of increasing [O/H] abundances with {\it decreasing} 
\tef, which is not congruous with expectations for open cluster stars nor 
predicted by NLTE calculations.  In fact, NLTE corrections for MS dwarfs with 
$T_{\mathrm{eff}} \la 5500 \; \mathrm{K}$ are predicted to be less than 0.05 dex 
and {\it decrease} with {\it decreasing} \teff according to 
\citet{2003A&A...402..343T}.

The triplet is highly sensitive to \tef, and thus the cool dwarf O abundance 
trend may be a consequence of the adopted \teff scale.  At $\sim 45 \; 
\mathrm{pc}$, the Hyades is the nearest open cluster and is well-studied 
photometrically and astrometrically, making the use of a color-temperature 
relation a seemingly reasonable choice for derivations of Hyades dwarf \tef.  
However, \citet{2004ApJ...600..946P} was unable to fit an isochrone to the 
photometry of the Hyades cluster, especially for the reddest stars, using several 
color-temperature relations from the literature.  They go on to demonstrate that 
their empirically calibrated luminosity-based temperature scale is in better 
agreement with the spectroscopically derived temperatures of \citet{paulson} and 
argue the mismatch between photometry and theoretical colors is due to systematic 
errors in the color-temperature relations.  In order to test the sensitivity of 
the cool dwarf abundances to our adopted photometric \teff scale, temperatures 
using the empirically calibrated Hyades isochrone of Pinsonneault et al. (2004; 
$T_{\mathrm{PTHS}}$) and the metallicity-dependent temperature-color calibration 
of Ram{\'i}rez \& Mel{\'e}ndez (2005; $T_{\mathrm{IRFM}}$) have been derived.  
The temperature scale of Ram{\'i}rez \&  Mel{\'e}ndez is derived using the 
infrared flux method (IRFM; Blackwell et al. 1990) and is an extension of the 
work of Alonso, Arribas, \& Mart{\'i}nez-Roger (1996, 1999).  The B-V colors from 
Table 1 and a cluster metallicity of $\mathrm{[Fe/H]} = 0.13$ were used to derive 
$T_{\mathrm{IRFM}}$ from the calibration of Ram{\'i}rez \& Mel{\'e}ndez (we use 
the coefficients for the metallicity-dependent polynomial correction that are 
applicable for $-0.5 < \mathrm{[Fe/H]} < +0.5$).  

In Figure 4a, $T_{\mathrm{PTHS}}$ and $T_{\mathrm{IRFM}}$ are plotted against 
the adopted photometric \tef.  The Pinsonneault et al. isochrone-based
temperatures of all but seven stars are higher than our adopted ones; temperature 
differences on a star-by-star basis are $|T_{\mathrm{PTHS}} - T_{\mathrm{eff}}| 
\leq 152 \; \mathrm{K}$, with differences for about half of the sample being $< 
60 \; \mathrm{K}$.  Temperatures from the temperature-color calibration of 
Ram{\'i}rez \& Mel{\'e}ndez are in better agreement with \tef,
with star-by-star differences being $|T_{\mathrm{IRFM}} - T_{\mathrm{eff}}| \leq 
57 \; \mathrm{K}$, with all but nine having differences $\leq 30 \; \mathrm{K}$.  
For both $T_{\mathrm{PTHS}}$ and $T_{\mathrm{IRFM}}$, absolute differences with 
\teff are approximately less than or equal to the $1\sigma$ error \tef, and thus 
no large-scale differences are found.  Oxygen abundances have been rederived 
using the new temperatures, and the results are plotted along with the original 
values in Figure 4b.  The abundances in Figure 4b have been culled of duplicates; 
in cases of overlap, preferred abundances are taken from the data set following 
the order KPNO-02, KPNO-04, and McD-04.  The KPNO-02 abundances are those derived 
using the mean EWs, as described above.  In the intermediate \teff range of 
5450-6100 K, the abundances of the dwarfs derived using $T_{\mathrm{PTHS}}$ are 
on average 0.06 dex lower than the original abundances.  Outside of this 
intermediate \teff range, the trends of increasing O abundances among both the 
warm and cool dwarfs remain.  The abundances derived using $T_{\mathrm{IRFM}}$ do 
not differ significantly from the original abundances, and again, the cool dwarf 
abundance trend persists.  In both cases, the differences in the abundances of 
the coolest stars are of the greatest magnitude, but they are generally within 
the calculated uncertainties.  {\it The pertinent result of this exercise is that 
the triplet abundance trend among the cool Hyades dwarfs presented here is
apparently not due to an inaccurate \teff scale}.  A similar conclusion was 
reached by SKHP in their study of the triplet in Pleiades dwarfs.  This result is 
not surprising given the abundance sensitivities in Table 5.  For the coolest 
stars in the sample, a \teff change of over 400 K would be required to bring 
their abundances into concordance with those of the intermediate \teff stars, 
where there is good star-to-star agreement.  Comparing the adopted photometric 
\teff scale to those of Pinsonneault et al. and of Ram{\'i}rez \& Mel{\'e}ndez 
suggests that uncertainties in our temperatures may be as high as 150 K, which is 
significantly less than what is required to eliminate the cool dwarf abundance 
trend. 
\marginpar{Fig.~4}

The morphology of the Hyades cool dwarf ($T_{\mathrm{eff}} \leq 5450 \; 
\mathrm{K}$) abundances is qualitatively similar to that recently seen in 
Pleiades dwarfs (SKHP), but significant differences do exist.  The Hyades [O/H] 
abundances sans duplicates and the SKHP Pleiades [O/H] abundances (derived using 
Kurucz models with no convective overshoot) are plotted in Figure 5.  The typical 
uncertainty in the Pleiades abundances is $\pm 0.08$ dex.  Although there are 
more Hyades stars with $T_{\mathrm{eff}} > 5200 \; \mathrm{K}$, no difference in 
the triplet abundances of the two clusters in this \teff region is discernible.  
This in itself is striking given that Hyades stars are metal-rich compared to 
Pleiades ($\mathrm{[Fe/H]} \approx 0.00$; Boesgaard \& Friel 1990) stars, 
suggesting the Pleiades [O/Fe] ratio could be as much as 0.13 dex higher than the 
Hyades.  It has been demonstrated that even a moderate, non-zero [O/Fe] ratio can 
have a significant effect on pre-main sequence (PMS) Li depletion predictions 
(Swenson et al. 1994; Piau \& Turck-Chi{\` e}ze 2002), and because of the 
prominent roles the Pleiades and Hyades play in the study of Li depletion 
mechanisms (e.g., Ford, Jeffries, \& Smalley 2002), the potentially larger [O/Fe] 
ratio for the Pleiades relative to the Hyades must be taken into consideration.
\marginpar{Fig.~5}

For stars with $T_{\mathrm{eff}} < 5200 \; \mathrm{K}$, the [O/H] abundances of
the two clusters clearly diverge, with the Hyades trend being less steep than 
that for the Pleiades.  The Hyades [O/H] abundances were fit with a quadratic
function, and the reduced $\chi^2$ between this fit and the Pleiades data was 
calculated.  Based on this, the Pleiades abundances deviate from those of the 
Hyades at a confidence level of $>99.999\%$.  In an attempt to explain this 
divergence and, as a result, shed light on the nature of the increasing triplet 
abundances with decreasing \teff, one naturally turns to the two possibly 
relevant physical characteristics that are disparate between these two clusters: 
metallicity and age.  A similar suggestion is offered by \citet{ks05} who 
compared their triplet results of six UMa dwarfs to the SKHP Pleiades abundances 
and found the UMa trend to be shallower than the Pleiades.  The UMa moving group 
is characterized by a sub-solar Fe abundance of $\mathrm{[Fe/H]} = -0.09$ 
\citep{1990ApJ...351..467B} and an age that is approximately coeval to the 
Hyades \citep{ks05}; UMa is an excellent sample with respect to the Hyades and 
Pleiades results to perform an empirical test of the sensitivity of the [O/H] 
trends to metallicity and age.  Accordingly, the King \& Schuler [O/H] UMa 
abundances are included in Figure 5.  In order to put the UMa abundances on a 
scale similar to the Hyades and Pleiades, the abundances have been increased by 
a constant to bring the abundance of the warmest UMa star into concordance with 
the mean abundance for Hyads in the range of $5450 \leq T_{\mathrm{eff}} \leq 
6000 \; \mathrm{K}$, the temperature range in which the two warmest UMa stars 
are located.  If metallicity is a factor, then an UMa trend that is steeper than 
the Pleiades trend would be expected.  Despite the paucity of UMa data points, 
this is clearly not seen.  Indeed, for stars with $T_{\mathrm{eff}} \leq 5300 \; 
\mathrm{K}$, the UMa abundances map well with the Hyades trend.  These limited 
data, while not conclusive, instead point to an age-related effect influencing 
the triplet abundance trends in cool cluster dwarfs.  We now turn our attention 
to two age-related phenomena previously suggested to possibly affect triplet 
abundances: chromospheric activity and photospheric spots.

\subsection{Chromospheric Activity}
The effects of a global chromosphere on the formation of spectral lines are not 
well understood.  There is some evidence that the large spread in Li abundances 
among Pleiades K-dwarfs is correlated with chromospheric activity indicators 
(King \& Schuler 2004; Soderblom et al. 1993), but it is far from clear whether 
the correlation is causal.  With respect to O, \citet{1995PASJ...47..463T} 
found appreciable increases in the modeled line strengths of the triplet in the 
solar spectrum when using Kurucz atmospheres modified to include a chromosphere. 
The increase is apparently due to the temperature rise at depths $\tau_{5000} 
\lesssim 10^{-4}$ and to the large fraction of O atoms that remain neutral. 
SKHP compared the O abundances derived from the triplet for the Pleiades and M34
to $H\alpha$ and \ion{Ca}{2} triplet chromospheric activity measures from the
literature and found no correlation.  A similar analysis for the Hyades sample 
is presented in Figure 6, where our [O/H] abundances are plotted against
\ion{Ca}{2} H+K activity indicators \citep{2002AJ....124..572P} and against
differences in observed and \tef-dependent mean values ($\Delta 
R'_{\mathrm{HK}}$).  No statistically significant correlation exists for either 
comparison.  Thus, the combined Hyades and Pleiades results suggest a global 
chromosphere does not contribute to triplet trends among cool cluster dwarfs.
\marginpar{Fig.~6}

More recently, \citet{2004A&A...423..677M} analyzed the triplet and [\ion{O}{1}]
$\lambda 6300$ forbidden line in spectra of 14 single-lined, 
chromospherically active RS CVn binaries, and in Pleiades, Hyades, and field 
dwarf, as well as in field giant, data from the literature.  They report 
increasingly discrepant results between the two indicators with higher
chromospheric activity levels, with the discrepancy primarily due to increasing
triplet abundances (see Figure 1 therein).  In general, the inconsonant results
are restricted to the RS CVn binaries and the Pleiades dwarfs- the other stellar 
subsets generally have lower levels of chromospheric activity and do not evince 
a triplet-activity correlation.  With regards to the Pleiades dwarfs (the [O/H]
and color-based \teff data for which are taken from SKHP), a correlation between 
triplet [O/Fe] abundances and X-ray-based activity indicators ($R_X$- ratio of 
X-ray and bolometric luminosities) at a confidence level of $>98\%$ is 
reported.  This should be viewed with caution due to the correlation (at the 
$\sim 97\%$ confidence level according to the linear correlation coefficient; 
data taken from Table 2 of Morel \& Micela 2004) between the Pleiades $R_X$ and 
\tef, and to the fact that the [O/Fe] abundances are highly correlated with 
\teff ($>99.9\%$).  The degeneracy of these correlations prevents firm 
conclusions from being made, and it is possible that the [O/Fe]-$R_X$ activity 
relation might be due to other \tef-related effects.  Indeed, Morel \& Micela 
state the Pleiades [O/H]-\teff relation found by SKHP may be masking an activity 
effect given the tight correlation between $(B-V)_0$-$R_X$ as reported by 
\citet{1999A&A...341..751M}; however, the converse argument is equally valid, 
i.e., the [O/Fe]-activity relation is masking a \teff effect.  Because no 
correlation between Pleiades triplet [O/H] abundances and $H\alpha$ and 
\ion{Ca}{2} chromospheric activity indicators exists, we find the data 
supporting a triplet-activity relation for the Pleiades unconvincing.

For the RS CVn binaries, the interpretation of the Morel \& Micela results and
their relevance to the current study are not clear.  The triplet [O/Fe]
abundances of the Rs CVn stellar subset show a statistically significant 
correlation at the $\sim 96\%$ confidence level with \ion{Ca}{2} H+K activity 
indicators, but a correlation with $R_X$ indicators is totally absent.  On the 
other hand, both activity indicators are not correlated with \teff, but the 
[O/Fe] abundances are correlated with \teff at a $>99.9\%$ confidence level.  
These relations do not provide a definitive picture of the connection between 
activity and the triplet O abundances for these stars, especially when the 
nature and inhomogeneity of the sample are considered.  RS CVn systems have 
enhanced chromospheric and photospheric (spots) activity, presumably due to 
effects resulting from interactions between the binary components.  The primary 
components are generally evolved stars, and the Morel \& Micela sample is 
composed mostly of sub-giants, according to the derived surface gravities (Morel 
et al. 2003; Morel et al. 2004).  Thus, comparing the RS CVn triplet results to 
those of the Hyades and Pleiades dwarfs may be unwarranted.

\subsection{Photospheric Spots}
The effect of photospheric spots on abundance derivations has received a healthy 
amount of attention in the literature, especially with respect to Li abundances 
as measured from the $\lambda 6707$ resonance feature (e.g., Xiong \& Deng 2005; 
Ford et al. 2002).  Unfortunately, such analyses are limited to speculation 
because spots and their areal coverage of stellar surfaces are not well 
understood.  The advent of doppler imaging has permitted the mapping of stellar 
surfaces and confirmed the presence of spots on target stars, but the 
technique is limited to stars with high rotational velocities ($40 \leq v \sin 
i \leq 80 \; \mathrm{km \; s}^{-1}$; Vogt \& Penrod 1983).  In order to reduce the 
rotationally-induced broadening of spectral features, stars that are typically 
included in abundance analyses are slow rotators and thus are not suitable for 
doppler imaging studies. 

Nonetheless, we have undertaken a plausibility study to investigate whether the 
presence of spots could produce the anomalous triplet O abundances observed 
among our cool Hyades dwarf sample.  We have utilized ``toy'' model atmospheres
that include flux contributions to the triplet region from both cool 
and hot spots, as well as from the quiescent star.  Our modeling scheme is a 
gross oversimplification of a complex interplay of atmospheric physics (i.e.,
magnetic fields, etc.) and temperature inhomogeneities, but it is useful as a 
first approximation of spot effects.  The first criterion to which our 
multi-component toy models adhere is the conservation of total luminosity, 
similar to that described by \citet{2002A&A...391..253F}, by using a weighted 
Stefan-Boltzmann relation:
\begin{displaymath}
T_{\mathrm{eff}} = a_{\mathrm{cool}}T^4_{\mathrm{cool}} + 
a_{\mathrm{hot}}T^4_{\mathrm{hot}} + a_{\mathrm{star}}T^4_{\mathrm{star}}
\end{displaymath}
where \teff is the adopted effective temperature for the star (Table 1), 
$a_{\mathrm{cool}}$, $a_{\mathrm{hot}}$, and $a_{\mathrm{star}}$ are the areal
coverages of the cool spots, hot spots, and quiescent star, respectively, and 
$T_{\mathrm{cool}}$, $T_{\mathrm{hot}}$, and $T_{\mathrm{star}}$ are the
corresponding temperatures.  The sum of the areal coverages are constrained such
that $$a_{\mathrm{cool}} + a_{\mathrm{hot}} + a_{\mathrm{star}} = 1.$$  Areal
coverages of $a_{\mathrm{cool}} = a_{\mathrm{hot}} = 0.20$ and initial spot
temperatures were chosen arbitrarily, and the Stefan-Boltzmann relation was then 
solved for $T_{\mathrm{star}}$.  Three model atmospheres characterized by either 
$T_{\mathrm{cool}}$, $T_{\mathrm{hot}}$, or $T_{\mathrm{star}}$ were constructed;
surface gravities and microturbulent velocities remain unchanged for the three
models for each star and are as they appear in Table 1.  With an
input O abundance equal to the mean abundance of Hyads in the range of $5450 
\leq T_{\mathrm{eff}} \leq 6100 \; \mathrm{K}$ ($\mathrm{[O/H]} = +0.25$), cool, 
hot, and star component spectra were synthesized using {\sf MOOG}.  Before 
combining the spectra into a composite spectrum, the contribution of flux at 
7774  {\AA}- approximately at the center of the \ion{O}{1} triplet- from the
three temperature components was determined by solving the Planck function for 
each.  The resulting values were then normalized to create a ``Planck factor''.  
Each component spectrum was multiplied by its areal coverage and Planck factor, 
and finally the components were added to achieve a composite spectrum.  The 
cool and hot temperatures were altered and the procedure iterated until the 
measured EWs of the composite spectrum were approximately equal (typically to 
within the measured EW uncertainties) to the observed EWs.

A sample of the component and composite spectra of HIP 18322 is presented in 
Figure 7, and the results of this exercise for three Hyads are given in Table 7. 
Assuming reasonable spot coverages and temperatures given in Table 7, the toy 
model employed here is able to satisfactorily reproduce the observed line 
strengths for all three stars.  It is evident from Figure 7 that if spots affect 
observed triplet line strengths, the dominant contribution comes from hot spots, 
i.e., faculae and/or plages; the cool spot regions have little or no effect.  
This provides significant freedom in choosing the spot coverages and 
temperatures for the present case, and it excludes uniqueness for our models.  A 
more sophisticated approach to this problem could include other spectral 
features, including molecular lines, with differing \teff sensitivities in order 
to constrain the areal coverages and temperatures of the models, assuming there 
are other features that show \teff relations similar to the triplet.  There are 
existing data suggesting this to be the case (e.g., SKHP).  While this exercise 
is in no way conclusive, we have been able to demonstrate that photospheric 
temperature inhomogeneities can possibly account for the high triplet abundances 
of the cool Hyades dwarfs presented here and thus possibly produce the triplet 
abundance trends observed in cool open cluster dwarfs.
\marginpar{Fig.~7}
\marginpar{Tab.~6}

\section{SUMMARY AND CLOSING REMARKS}
Oxygen abundances have been derived under the assumption of LTE from the 
high-excitation \trip in high-resolution spectra for 45 Hyades dwarfs in the 
effective temperature range of $4300 - 6300 \; \mathrm{K}$.  The warmest stars 
in the sample ($T_{\mathrm{eff}} > 6200 \; \mathrm{K}$) show evidence of 
increasing triplet abundances with increasing \tef, as predicted by current NLTE 
calculations.  A striking increase of triplet abundances with {\it decreasing} 
\teff is observed for stars with $T_{\mathrm{eff}} < 5450 \; \mathrm{K}$, 
contrary to expectations and to NLTE predictions.  Triplet abundances have been
rederived using two additional temperature scales, and the morphology of the cool 
dwarf abundance trend is unaltered, suggesting the trend is not due to erroneous 
\tef.  Comparing the Hyades trend to 
that previously observed among Pleiades dwarfs (SKHP) reveals both similarities 
and differences.  At \teff greater than $\sim 5200 \; \mathrm{K}$, no difference 
in the abundances at a given \teff for the two clusters is discernible, raising 
the possibility of a larger [O/Fe] ratio for the Pleiades relative to the Hyades. 
For stars with \teff less than $\sim 5200 \; \mathrm{K}$, the two trends diverge 
with the Hyades trend being less steep than that for the Pleiades.  Another 
difference is the star-to-star abundance scatter seen for Pleiads with 
$T_{\mathrm{eff}} < 5300 \; \mathrm{K}$ is not seen among the Hyads.  

Recently derived triplet abundances for six dwarfs that are members of the Ursa 
Major moving group are compared to the Hyades and Pleiades results.  UMa members 
have a sub-solar metallicity ($\mathrm{[Fe/H]}=-0.09$; Boesgaard \& Friel 1990), 
and an age that is approximately coeval with that of the Hyades \citep{ks05}.  
Despite the small number of UMa stars, their abundances closely 
follow the Hyades results, suggesting an age-related effect, as opposed to 
metallicity, is a material factor in triplet O abundance trends in cool cluster 
dwarfs.  No correlation is found between \ion{Ca}{2} H+K activity chromospheric 
activity indicators and Hyades [O/H] abundances.  SKHP report a similar lack of
correlation between $H\alpha$ and \ion{Ca}{2} triplet chromospheric activity 
indicators and Pleiades triplet O abundances.  Thus, we are unable to find 
evidence for an activity-triplet relation as suggested by the theoretical results
of \citet{1995PASJ...47..463T}.  On the other hand, we demonstrate to first 
order that photospheric temperature inhomogeneities possibly due to spots, 
faculae, and/or plages can plausibly produce anomalously high triplet 
abundances.  A simple toy model consisting of arbitrarily chosen line flux 
contributions from cool and hot spots is able to reproduce observed equivalent 
widths for three Hyads with differing \teff and triplet abundances.  The results 
do not conclusively point to spots as the source of the triplet trends in cool 
cluster dwarfs, but they are provocative and require further investigation.

There is concordance among the triplet O abundances for stars in the intermediate
\teff range of 5450-6100 K.  Over this \teff range, the NLTE corrections of 
\citet{2003A&A...402..343T} differ by $\leq 0.05$ dex, and thus the abundances given 
relative to solar values are essentially free of NLTE-related anomalies.  The mean 
abundance for this group of 15 stars (excluding HIP 15310 which has questionable 
cluster membership) is 
$\mathrm{[O/H]} = +0.25 \pm 0.02$, where the quoted error is the uncertainty in 
the mean.  This value is in good agreement with that found by \citet{kphd}, who 
derived an LTE abundance using the triplet of $\mathrm{[O/H]} = +0.22 \pm 
0.04$\footnotemark[7].  \citet{1993ApJ...412..173G} performed an NLTE analysis 
of the triplet in 25 Hyades F dwarfs and reported a mean O abundance of $-0.06 
\pm 0.15$.  This value is appreciably lower than that of the present study and 
of \citet{kphd}.  The difference in the Garc{\'i}a L{\'o}pez et al. and the King 
1993 results is discussed thoroughly in \citet{1996AJ....112.2650K}, who come to 
the conclusion that the bulk of the difference is most likely due to the NLTE 
corrections adopted by Garc{\'i}a L{\'o}pez et al. and to different solar EWs 
adopted by the two studies.  Finally, \citet{1996AJ....112.2650K} derived a 
Hyades O abundance from two dwarfs using spectral synthesis of the 
$\lambda 6300$ [\ion{O}{1}] feature and find a mean abundance of 
$\mathrm{[O/H]} = +0.15 \pm 0.01$.  Comparing this result with that of the
triplet in the intermediate group of stars here reveals a $\sim 0.10$
dex difference, possibly indicating the presence of underestimated internal 
errors or unaccounted for systematic errors in one or both studies.  For 
instance, if the luminosity-based \teff scale from the Hyades isochrone of 
\citet{2004ApJ...600..946P} is used to derive the triplet abundances of the 
intermediate group of stars, the mean abundance reduces to $\mathrm{[O/H]} 
= +0.19 \pm 0.02$.  If we take the [\ion{O}{1}]-based and triplet-based O 
abundances as lower and upper limits, respectively, it appears the Hyades O 
abundance falls in the range of $0.15 \lesssim \mathrm{[O/H]} \lesssim 0.25 \; 
\mathrm{dex}$.

\footnotetext[7]{The quoted abundance from King 1993 is that using an updated 
measurement of the EW for one of his program stars, as described by King \& 
Hiltgen 1996}

\acknowledgements
S.C.S. thanks the South Carolina Space Grant Consortium for providing support
through the Graduate Student Research Fellowship.  S.C.S. and J.R.K. gratefully 
acknowledge support for this work by grants AST 00-86576 and AST 02-39518 to 
J.R.K. from the National Science Foundation.  D.M.T. and M.H.P. thank the National
Science Foundation for support from grants AST-0205789 and AST-0206008 to the 
Ohio State University Research Foundation.  We also thank Ms. Abigail Daane, 
Mr. Roggie Boone, and Ms. Angela Hansen for their assistance with the McD-04, 
KPNO-04, and KPNO-02 observations.  This research has made use of the SIMBAD 
database, operated at CDS, Strasbourg, France and NASA's Astrophysics Data 
System Bibliographic Services.  We would like to acknowledge the anonymous
referee, whose comments have led to an improved paper.

\newpage
%References:

\clearpage
%Figures:
\begin{figure}
\plotone{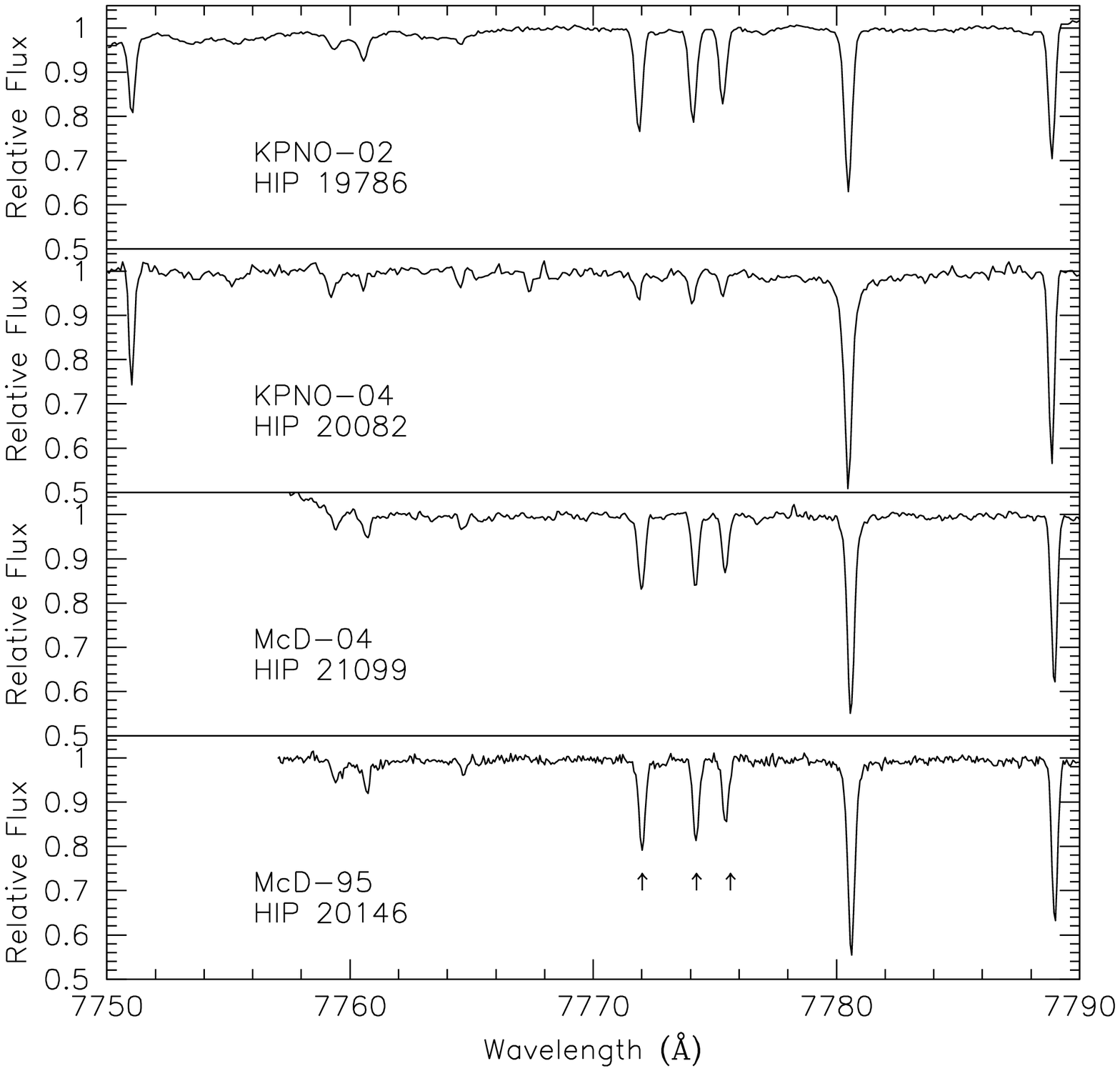}
\caption{Sample high-resolution spectra of Hyades dwarfs from the Kitt Peak 
National Observatory 4-m and  the McDonald Observatory 2.7-m telescopes.  The 
source of each spectrum, as described in the text, is given, and the \trip is 
marked in the lower panel.  Only the two reddest lines were measured for HIP 
20082 ($T_{\mathrm{eff}} = 4784 \; \mathrm{K}$).}
\end{figure}

\begin{figure}
\plotone{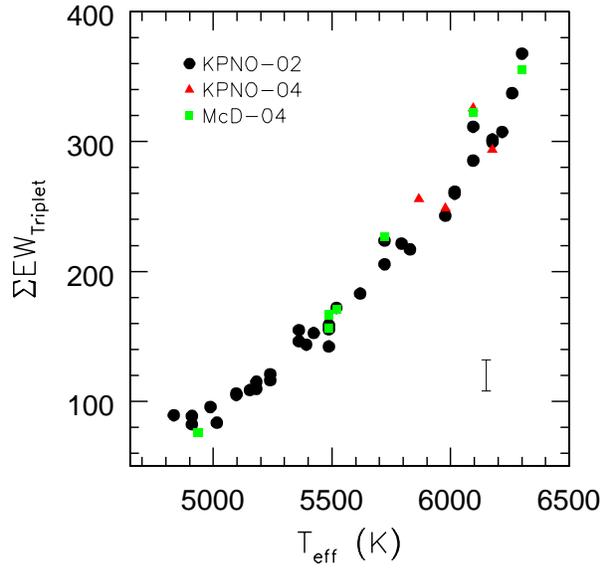}
\caption{Sum of the triplet EWs as a function of \teff for the combined Hyades
data set.  The points are distinguished by the data set from which they are 
measured.  The vertical bar represents the $1\sigma$ uncertainty in the 
combined EWs.  Only stars for which all three lines of the \trip are measurable 
are included.}
\end{figure}

\begin{figure}
\plotone{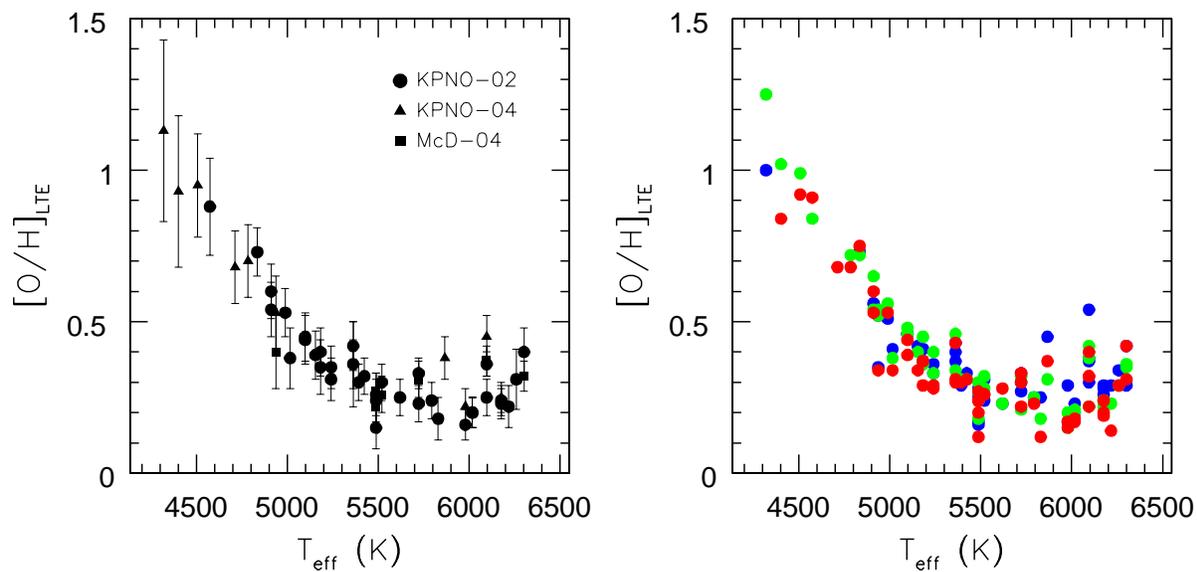}
\caption{Left- Relative LTE O abundances vs. \teff for the combined Hyades data 
set.  The points are again distinguished by the data set from which they are 
derived.  The error bars represent the total internal abundance uncertainties. 
Right- Line-by-line relative LTE O abundances vs. \teff for the combined Hyades
data set.  Abundances derived from the $\lambda7772$ line are given in blue, 
$\lambda7774$ are green, and $\lambda7775$ are red.}
\end{figure}

\begin{figure}
\plotone{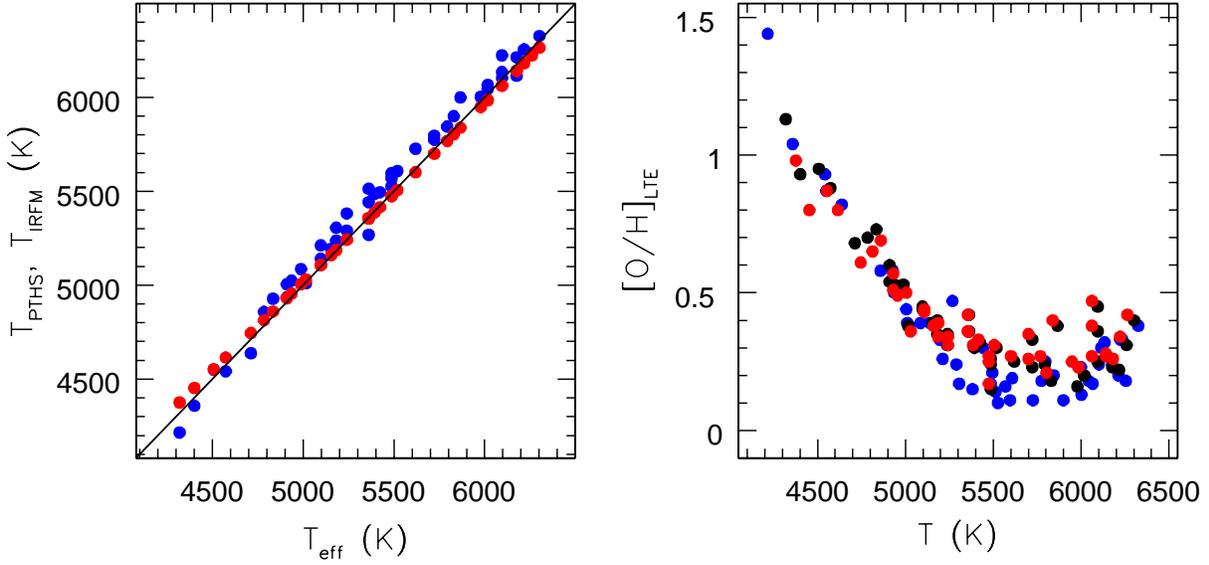}
\caption{(a) Temperatures from the isochrone of Pinsonneualt et al. (2004; blue) 
and from the metallicity-dependent temperature-color calibration of Ram{\'i}rez 
\& Mel{\'e}ndez (2005; red) vs. the adopted photometry-based temperatures.  The
diagonal line has a slope of unity and is the line of equality.  (b) Relative LTE
O abundances vs. temperature.  Abundances derived using the adopted 
photometry-based temperatures (\tef) are plotted in black, the Pinsonneault et 
al. isochrone-base temperatures ($T_{\mathrm{PTHS}}$) in blue, and the 
Ram{\'i}rez \& Mel{\'e}ndez calibration-based  temperatures ($T_{\mathrm{IRFM}}$) 
in red.  The sample has been trimmed of duplicates as described in the text.}
\end{figure}

\begin{figure}
\plotone{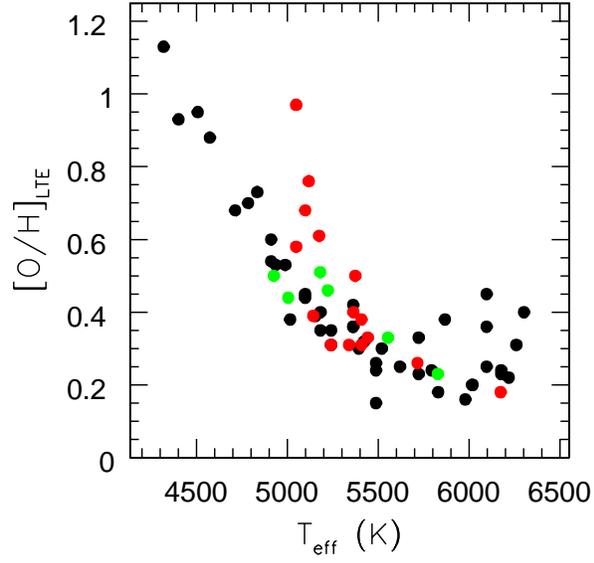}
\caption{Hyades (black), Pleiades (red), and UMa (green) relative LTE O 
abundances vs. \tef.  The Hyades sample is trimmed of duplicates as described in 
the text.  The Pleiades data are from SKHP, and the UMa data are from 
\citet{ks05}}
\end{figure}

\begin{figure}
\plotone{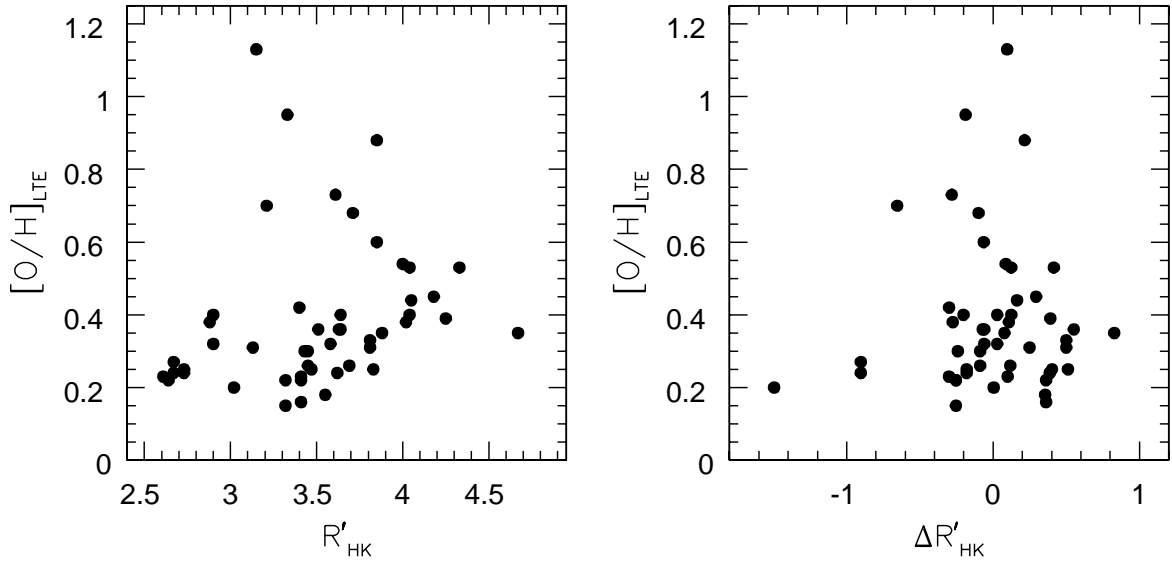}
\caption{Left- Hyades relative LTE O abundances vs. \ion{Ca}{2} H+K activity 
indicators.  Activity data are from \citet{2002AJ....124..572P}.  Right- Hyades
relative LTE O abundances vs. \ion{Ca}{2} H+K activity residuals.  The residuals are
the difference in the observed and \tef-dependent fitted values.}
\end{figure}

\begin{figure}
\plotone{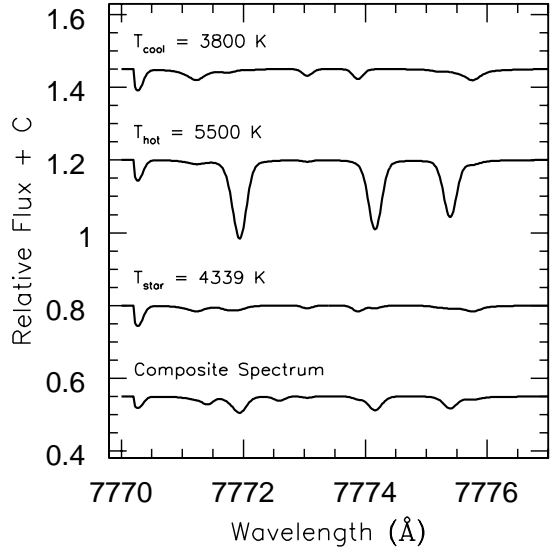}
\caption{Synthesized component and composite spectra of HIP18322 used in the spot
plausibility study.  The spectra are synthesized with input O abundance of 
$\mathrm{[O/H]} = +0.25$, the mean abundance of Hyads in the range $5450 
\leq T_{\mathrm{eff}} \leq 6100 \; \mathrm{K}$.}
\end{figure}

\clearpage
%Tables:
\begin{deluxetable}{lrrrcccccccl}
\rotate
\tablecolumns{12}
\tablenum{1}
\tablewidth{0pt}
\tablecaption{Hyades Dwarfs Parameters}
\tablehead{
     \colhead{}&
     \colhead{}&
     \colhead{}&
     \colhead{}&
     \colhead{($B-V$)}&
     \colhead{}&
     \colhead{$T_{\mathrm{eff}}$}&
     \colhead{}&
     \colhead{$\log g$}&
     \colhead{}&
     \colhead{$\xi$}&
     \colhead{}\\
     \colhead{HIP}&
     \colhead{}&
     \colhead{HD}&
     \colhead{}&
     \colhead{(Mag)}&
     \colhead{Comment\tablenotemark{a}}&
     \colhead{(K)}&
     \colhead{}&
     \colhead{(cgs)}&
     \colhead{}&
     \colhead{(km $\mathrm{s}^{-1}$)}&
     \colhead{Telescope\tablenotemark{b}}
     }
\startdata
13806 && \nodata&& 0.86 & P & 5097 && 4.60 && 1.10 & KPNO-02\\
13976 &&  18632 && 0.93 & P & 4910 && 4.63 && 1.04 & KPNO-02\\
14976 &&  19902 && 0.73 & A & 5487 && 4.54 && 1.24 & KPNO-02,McD-04\\
15310 &&  20439 && 0.62 & A & 5866 && 4.47 && 1.40 & KPNO-04\\
16529 && \nodata&& 0.84 & A & 5154 && 4.59 && 1.12 & KPNO-02\\
16908 && \nodata&& 0.92 & Y & 4936 && 4.63 && 1.05 & KPNO-04,McD-04\\
18322 && 286363 && 1.07 & Y & 4573 && 4.68 && 0.94 & KPNO-02\\
18327 && 285252 && 0.90 & P & 4988 && 4.62 && 1.07 & KPNO-02\\
18946 && 265348 && 1.10 & Y & 4507 && 4.69 && 0.92 & KPNO-04\\
19098 && 285367 && 0.89 & A & 5015 && 4.61 && 1.08 & KPNO-02\\
19148 &&  25825 && 0.59 & A & 5978 && 4.44 && 1.45 & KPNO-02,KPNO-04\\
19263 && 285482 && 1.01 & Y & 4712 && 4.66 && 0.98 & KPNO-04\\
19441 && \nodata&& 1.19 & Y & 4319 && 4.72 && 0.87 & KPNO-04\\
19781 &&  26756 && 0.69 & A & 5619 && 4.51 && 1.30 & KPNO-02\\
19786 &&  26767 && 0.64 & A & 5793 && 4.48 && 1.37 & KPNO-02\\
19793 &&  26736 && 0.66 & A & 5722 && 4.49 && 1.34 & KPNO-02,McD-04\\
19796 &&  26784 && 0.51 & A & 6301 && 4.37 && 1.61 & KPNO-02,McD-04\\
19934 && 284253 && 0.81 & A & 5240 && 4.58 && 1.15 & KPNO-02\\
20082 && 285690 && 0.98 & P & 4784 && 4.65 && 1.00 & KPNO-04\\
20130 &&  27250 && 0.75 & A & 5423 && 4.55 && 1.22 & KPNO-02\\
20146 &&  27282 && 0.72 & A & 5519 && 4.53 && 1.26 & KPNO-02,McD-95\\
20237 &&  27406 && 0.56 & A & 6095 && 4.42 && 1.51 & KPNO-02\\
20480 &&  27732 && 0.76 & A & 5392 && 4.55 && 1.21 & KPNO-02\\
20492 &&  27771 && 0.86 & A & 5097 && 4.60 && 1.10 & KPNO-02\\
20557 &&  27808 && 0.52 & A & 6259 && 4.38 && 1.59 & KPNO-02\\
20712 &&  28033 && 0.56 & A & 6095 && 4.42 && 1.51 & KPNO-04,McD-04\\
20741 &&  28099 && 0.66 & A & 5722 && 4.49 && 1.34 & KPNO-02\\
20762 &&  32347 && 1.15 & Y & 4401 && 4.70 && 0.89 & KPNO-04\\
20815 &&  28205 && 0.54 & A & 6176 && 4.40 && 1.55 & KPNO-02\\
20826 &&  28237 && 0.56 & A & 6095 && 4.42 && 1.51 & KPNO-02\\
20827 && 285830 && 0.93 & P & 4910 && 4.63 && 1.04 & KPNO-02\\
20949 && 283704 && 0.77 & A & 5361 && 4.56 && 1.20 & KPNO-02\\
20951 && 285773 && 0.83 & P & 5182 && 4.59 && 1.13 & KPNO-02\\
21099 &&  28593 && 0.73 & A & 5487 && 4.54 && 1.24 & KPNO-02,McD-04\\
21112 &&  28635 && 0.54 & A & 6176 && 4.40 && 1.55 & KPNO-02,KPNO-04\\
21317 &&  28992 && 0.63 & A & 5829 && 4.47 && 1.39 & KPNO-02\\
21637 &&  29419 && 0.58 & A & 6017 && 4.43 && 1.47 & KPNO-02\\
21741 && 284574 && 0.81 & A & 5240 && 4.58 && 1.15 & KPNO-02\\
22380 &&  30505 && 0.83 & A & 5182 && 4.59 && 1.13 & KPNO-02\\
22422 &&  30589 && 0.58 & A & 6017 && 4.43 && 1.47 & KPNO-02\\
22566 &&  30809 && 0.53 & A & 6217 && 4.39 && 1.57 & KPNO-02\\
23312 && \nodata&& 0.96 & P & 4834 && 4.64 && 1.02 & KPNO-02\\
23498 && \nodata&& 0.77 & A & 5361 && 4.56 && 1.20 & KPNO-04\\
23750 && 240648 && 0.73 & A & 5487 && 4.54 && 1.24 & KPNO-02\\
24923 && 242780 && 0.77 & A & 5361 && 4.56 && 1.20 & KPNO-02\\
      &&        &&      &   &	   &&	   &&	   &       \\
Sun   &&        &&      &   & 5777 && 4.44 && 1.38 & KPNO-02,KPNO-04,McD-04\\
\enddata

\tablenotetext{a}{Provides the source of the $(B-V)$ values: A- Allende Prieto
                  \& Lambert 1999; P- Paulson et al. 2003; Y- Yong et al. 2004}
\tablenotetext{b}{Denotes with which telescope the object was observed, as 
                  described in \textsection 2: KPNO-02- KPNO 4-m 2002; KPNO-04- 
		  KPNO 4-m 2004; McD-95- McDonald Observatory 2.7-m 1995; 
		  McD-04- McDonald Observatory 2.7-m 2004} 
\end{deluxetable}

\begin{deluxetable}{lrcrrrccrrrcrrr}
\tablecolumns{15}
\tablewidth{0pt}
\rotate
\tablenum{2}
\tablecaption{Triplet Equivalent Widths: KPNO-02}
\tablehead{
     \colhead{}&
     \colhead{}&
     \multicolumn{4}{c}{Blue Order}&
     \colhead{}&
     \multicolumn{4}{c}{Red Order}&
     \colhead{}&
     \multicolumn{3}{c}{Mean}\\
     \cline{3-6} \cline{8-11} \cline{13-15}\\
     \colhead{HIP}&
     \colhead{}&
     \colhead{S/N}&
     \colhead{$\mathrm{EW}_{7772}$}&
     \colhead{$\mathrm{EW}_{7774}$}&
     \colhead{$\mathrm{EW}_{7775}$}&
     \colhead{}&
     \colhead{S/N}&
     \colhead{$\mathrm{EW}_{7772}$}&
     \colhead{$\mathrm{EW}_{7774}$}&
     \colhead{$\mathrm{EW}_{7775}$}&
     \colhead{}&
     \colhead{$\mathrm{EW}_{7772}$}&
     \colhead{$\mathrm{EW}_{7774}$}&
     \colhead{$\mathrm{EW}_{7775}$}\\
     \colhead{}&
     \colhead{}&
     \colhead{}&
     \colhead{(m{\AA})}&
     \colhead{(m{\AA})}&
     \colhead{(m{\AA})}&
     \colhead{}&
     \colhead{}&
     \colhead{(m{\AA})}&
     \colhead{(m{\AA})}&
     \colhead{(m{\AA})}&
     \colhead{}&
     \colhead{(m{\AA})}&
     \colhead{(m{\AA})}&
     \colhead{(m{\AA})}
     }
\startdata
13806 && 290 &  38.1 &  36.0 &  29.7 && 194 &  44.5 &  36.5 &  27.2 &&  41.3 &  36.3 &  28.5\\
13976 && 490 &  30.7 &  30.0 &  21.9 && 320 &  34.4 &  25.9 &  21.6 &&  32.6 &  28.0 &  21.8\\
14976 && 340 &  57.8 &  54.5 &  42.6 && 260 &  65.2 &  51.2 &  39.5 &&  61.5 &  52.9 &  41.1\\
16529 && 265 &  41.4 &  40.1 &  28.8 && 200 &  45.8 &  34.4 &  26.7 &&  43.6 &  37.3 &  27.8\\
18322 && 245 &\nodata&  18.8 &  16.2 && 170 &\nodata&  18.4 &  15.5 &&\nodata&  18.6 &  15.9\\
18327 && 295 &  36.4 &  33.2 &  27.5 && 200 &  36.2 &  33.8 &  24.2 &&  36.3 &  33.5 &  25.9\\
19098 && 208 &  32.6 &  28.0 &  21.5 && 195 &  34.7 &  29.0 &  21.3 &&  33.7 &  28.5 &  21.4\\
19148 && 465 &  91.0 &  80.7 &  65.7 && 310 &  93.0 &  82.6 &  72.7 &&  92.0 &  81.7 &  69.2\\
19781 && 300 &  65.7 &  64.9 &  51.5 && 218 &  71.8 &  57.3 &  54.7 &&  68.8 &  61.1 &  53.1\\
19786 && 455 &  82.7 &  76.3 &  63.7 && 300 &  85.4 &  74.8 &  60.2 &&  84.1 &  75.6 &  62.0\\
19793 && 405 &  82.0 &  80.6 &  63.2 && 295 &  87.3 &  70.1 &  64.2 &&  84.7 &  75.4 &  63.7\\
19796 && 460 & 139.0 & 130.1 & 110.4 && 284 & 135.7 & 110.0 & 110.0 && 137.4 & 120.1 & 110.2\\
19934 && 293 &  44.4 &  39.6 &  32.6 && 190 &  46.5 &  40.9 &  28.5 &&  45.5 &  40.3 &  30.6\\
20130 && 286 &  56.5 &  53.6 &  39.9 && 208 &  61.6 &  49.5 &  44.0 &&  59.1 &  51.6 &  42.0\\
20146 && 289 &  61.8 &  61.5 &  46.5 && 194 &  71.1 &  57.5 &  45.6 &&  66.5 &  59.5 &  46.1\\
20237 && 390 & 115.8 & 111.1 &  86.8 && 300 & 116.8 & 104.4 &  87.9 && 116.3 & 107.8 &  87.4\\
20480 && 275 &  50.7 &  49.4 &  38.7 && 186 &  58.6 &  49.6 &  40.4 &&  54.7 &  49.5 &  39.6\\
20492 && 320 &  37.7 &  37.5 &  27.7 && 205 &  44.6 &  36.5 &  25.8 &&  41.2 &  37.0 &  26.8\\
20557 && 355 & 129.0 & 118.6 &  96.8 && 298 & 125.4 & 106.3 &  98.5 && 127.2 & 112.4 &  97.7\\
20741 && 340 &  76.2 &  74.0 &  56.5 && 240 &  84.1 &  62.8 &  57.4 &&  80.2 &  68.4 &  57.0\\
20815 && 387 & 114.8 & 106.6 &  88.5 && 260 & 112.4 &  97.4 &  79.8 && 113.6 & 102.0 &  84.2\\
20826 && 390 & 107.0 &  98.3 &  80.2 && 270 & 113.5 &  91.6 &  80.2 && 110.2 &  95.0 &  80.2\\
20827 && 285 &  30.5 &  32.9 &  24.7 && 190 &  35.4 &  30.9 &  23.0 &&  33.0 &  31.9 &  23.9\\
20949 && 238 &  55.4 &  56.9 &  40.0 && 180 &  57.1 &  53.3 &  47.0 &&  56.3 &  55.1 &  43.5\\
20951 && 320 &  41.6 &  43.6 &  29.6 && 200 &  44.4 &  39.5 &  31.5 &&  43.0 &  41.6 &  30.6\\
21099 && 325 &  53.7 &  47.1 &  39.9 && 220 &  55.5 &  52.4 &  35.7 &&  54.6 &  49.8 &  37.8\\
21112 && 347 & 112.7 & 105.5 &  84.0 && 255 & 117.1 &  97.8 &  85.9 && 114.9 & 101.7 &  85.0\\
21317 && 230 &  84.9 &  76.5 &  60.6 && 180 &  87.6 &  70.1 &  54.1 &&  86.3 &  73.3 &  57.4\\
21637 && 380 &  96.5 &  90.1 &  73.1 && 250 &  98.6 &  87.8 &  73.8 &&  97.6 &  89.0 &  73.5\\
21741 && 343 &  46.5 &  44.1 &  33.1 && 196 &  47.2 &  42.4 &  28.5 &&  46.9 &  43.3 &  30.8\\
22380 && 280 &  40.9 &  39.1 &  28.5 && 189 &  48.5 &  35.6 &  26.8 &&  44.7 &  37.4 &  27.7\\
22422 && 349 &  98.7 &  89.6 &  71.4 && 256 & 100.7 &  87.8 &  74.9 &&  99.7 &  88.7 &  73.2\\
22566 && 297 & 116.5 & 108.2 &  84.4 && 265 & 122.7 & 101.3 &  81.8 && 119.6 & 104.8 &  83.1\\
23312 && 256 &  30.1 &  30.9 &  24.5 && 162 &  39.6 &  29.2 &  24.4 &&  34.9 &  30.1 &  24.5\\
23750 && 261 &  58.0 &  59.1 &  45.6 && 176 &  61.4 &  52.8 &  40.5 &&  59.7 &  56.0 &  43.1\\
24923 && 274 &  56.7 &  48.6 &  38.1 && 163 &  63.6 &  48.5 &  37.0 &&  60.2 &  48.6 &  37.6\\
      &&     &       &       &       &&     &       &       &       &&       &       &      \\
Sun   && 342 &  64.3 &  61.0 &  50.3 && 219 &  68.8 &  58.7 &  47.5 &&  66.6 &  59.9 &  48.9\\
\enddata
\end{deluxetable}

\setcounter{page}{31}
\begin{deluxetable}{lrcrrr}
\tablecolumns{6}
\tablewidth{0pt}
\tablenum{3}
\tablecaption{Triplet Equivalent Widths: KPNO-04 \& McDonald}
\tablehead{
     \colhead{HIP}&
     \colhead{}&
     \colhead{S/N}&
     \colhead{$\mathrm{EW}_{7772}$}&
     \colhead{$\mathrm{EW}_{7774}$}&
     \colhead{$\mathrm{EW}_{7775}$}\\
     \colhead{}&
     \colhead{}&
     \colhead{}&
     \colhead{(m{\AA})}&
     \colhead{(m{\AA})}&
     \colhead{(m{\AA})}     
     }
\startdata
\cutinhead{KPNO-04}
15310 && 142 & 103.4 &  82.5 &  69.9\\
16908 && 210 &  31.8 &  28.2 &\nodata\\
18946 && 206 &\nodata&  18.2 &  11.9\\
19148 && 125 & 100.0 &  83.4 &  65.1\\
19263 && 188 &\nodata&\nodata&  14.9\\
19441 && 212 &  11.0 &  13.2 &\nodata\\
20082 && 221 &\nodata&  25.8 &  18.0\\
20712 && 123 & 130.4 & 108.0 &  87.4\\
20762 && 220 &\nodata&  12.8 &   6.9\\
21112 && 139 & 114.1 &  97.9 &  81.7\\
23498 &&  85 &  56.9 &\nodata&  34.7\\
      &&     &       &       &      \\
Sun   && 272 &  65.6 &  58.0 &  44.4\\
\cutinhead{McDonald}
14976 && 237 &  67.0 &  55.2 &  44.1\\
16908 && 131 &  29.8 &  28.7 &  17.6\\
19793 && 184 &  88.7 &  76.2 &  61.4\\
19796 && 208 & 133.0 & 121.5 & 100.7\\
20712 && 131 & 125.9 & 109.5 &  87.2\\
21099 && 190 &  60.3 &  53.4 &  42.6\\
      &&     &       &       &      \\
20146\tablenotemark{a} && 234 &  67.2 &  58.1 &  45.6\\
      &&     &       &       &      \\
Sun   && 825 &  72.8 &  60.2 &  48.2\\
\enddata

\tablenotetext{a}{Observed with the McDonald Observatory 2.7-m telescope in 1995}
\end{deluxetable}

\begin{deluxetable}{lrrr}
\tablecolumns{4}
\tablewidth{0pt}
\tablenum{4}
\tablecaption{Oxygen Triplet Abundances and Uncertainties}
\tablehead{
     \colhead{HIP}&
     \colhead{}&
     \colhead{[O/H]}&
     \colhead{$\sigma$}
     }

\startdata
\cutinhead{KPNO-02\tablenotemark{a}}
13806 && 0.45 & $\pm 0.08$\\
13976 && 0.54 & $\pm 0.09$\\
14976 && 0.24 & $\pm 0.07$\\
16529 && 0.39 & $\pm 0.08$\\
18322 && 0.88 & $\pm 0.16$\\
18327 && 0.53 & $\pm 0.08$\\
19098 && 0.38 & $\pm 0.10$\\
19148 && 0.16 & $\pm 0.05$\\
19781 && 0.25 & $\pm 0.06$\\
19786 && 0.24 & $\pm 0.06$\\
19793 && 0.33 & $\pm 0.05$\\
19796 && 0.40 & $\pm 0.08$\\
19934 && 0.31 & $\pm 0.07$\\
20130 && 0.32 & $\pm 0.06$\\
20146 && 0.30 & $\pm 0.06$\\
20237 && 0.36 & $\pm 0.04$\\
20480 && 0.30 & $\pm 0.06$\\
20492 && 0.44 & $\pm 0.08$\\
20557 && 0.31 & $\pm 0.10$\\
20741 && 0.23 & $\pm 0.06$\\
20815 && 0.23 & $\pm 0.05$\\
20826 && 0.25 & $\pm 0.06$\\
20827 && 0.60 & $\pm 0.09$\\
20949 && 0.42 & $\pm 0.08$\\
20951 && 0.40 & $\pm 0.08$\\
21099 && 0.15 & $\pm 0.07$\\
21112 && 0.24 & $\pm 0.05$\\
21317 && 0.18 & $\pm 0.07$\\
21637 && 0.20 & $\pm 0.05$\\
21741 && 0.35 & $\pm 0.07$\\
22380 && 0.35 & $\pm 0.09$\\
22422 && 0.20 & $\pm 0.05$\\
22566 && 0.22 & $\pm 0.07$\\
23312 && 0.73 & $\pm 0.08$\\
23750 && 0.26 & $\pm 0.07$\\
24923 && 0.36 & $\pm 0.08$\\
\cutinhead{KPNO-04}
15310 && 0.38 & $\pm 0.07$\\
16908 && 0.53 & $\pm 0.12$\\
18946 && 0.95 & $\pm 0.17$\\
19148 && 0.22 & $\pm 0.06$\\
19263 && 0.68 & $\pm 0.12$\\
19441 && 1.13 & $\pm 0.30$\\
20082 && 0.70 & $\pm 0.12$\\
20712 && 0.45 & $\pm 0.07$\\
20762 && 0.93 & $\pm 0.25$\\
21112 && 0.25 & $\pm 0.05$\\
23498 && 0.36 & $\pm 0.14$\\
\cutinhead{McDonald}
14976 && 0.27 &$\pm 0.06$\\
16908 && 0.40 &$\pm 0.12$\\
19793 && 0.31 &$\pm 0.06$\\
19796 && 0.32 &$\pm 0.05$\\
20712 && 0.37 &$\pm 0.05$\\
21099 && 0.22 &$\pm 0.06$\\
20146 && 0.26 &$\pm 0.06$\\
\enddata

\tablenotetext{a}{KPNO-02 abundances are those derived using the mean EWs.}

\end{deluxetable}

\begin{deluxetable}{lclcrrr}
\tablecolumns{7}
\tablewidth{0pt}
\tabletypesize{\footnotesize}
\tablenum{5}
\tablecaption{Line-by-line Triplet Abundance Sensitivities}
\tablehead{
     \colhead{}&
     \colhead{}&
     \colhead{}&
     \colhead{}&
     \multicolumn{3}{c}{Abundance Sensitivity}\\
     \cline{5-7}\\
     \colhead{HIP}&
     \colhead{}&
     \colhead{Parameter}&
     \colhead{}&
     \colhead{7772}&
     \colhead{7774}&
     \colhead{7775}
     }
\startdata
21112 ($T_{\mathrm{eff}} = 6176 \; \mathrm{K}$) && $\Delta T_{\mathrm{eff}} = \pm150 \; \mathrm{K}$             && $\mp 0.11$     & $\mp 0.12$     & $\mp 0.12$\\
      && $\Delta \log g = \pm 0.25 \; \mathrm{dex}$                   && $\pm 0.04$     & $\pm 0.06$     & $\pm 0.07$\\
      && $\Delta \xi = \pm 0.30 \; \mathrm{km} \; \mathrm{s}^{-1}$    && $\mp 0.04$     & $\mp 0.04$     & $\mp 0.04$\\
19786 ($T_{\mathrm{eff}} = 5793 \; \mathrm{K}$) && $\Delta T_{\mathrm{eff}} = \pm150 \; \mathrm{K}$             && $\mp 0.15$     & $\mp 0.15$     & $\mp 0.15$\\
      && $\Delta \log g = \pm 0.25 \; \mathrm{dex}$                   && $^{+0.08}_{-0.03}$ & $^{+0.09}_{-0.04}$ & $^{+0.09}_{-0.05}$\\
      && $\Delta \xi = \pm 0.30 \; \mathrm{km} \; \mathrm{s}^{-1}$    && $\mp 0.03    $ & $\mp 0.03    $ & $\mp 0.03$\\
20082 ($T_{\mathrm{eff}} = 4784 \; \mathrm{K}$) && $\Delta T_{\mathrm{eff}} = \pm150 \; \mathrm{K}$             && \nodata            & $^{-0.23}_{+0.27}$ & $^{-0.23}_{+0.27}$\\
      && $\Delta \log g = \pm 0.25 \; \mathrm{dex}$                   && \nodata            & $^{+0.05}_{-0.08}$ & $^{+0.06}_{-0.09}$\\
      && $\Delta \xi = \pm 0.30 \; \mathrm{km} \; \mathrm{s}^{-1}$    && \nodata        & $\mp 0.01$     & $\mp 0.01$\\
\enddata
\end{deluxetable}

\begin{deluxetable}{lrrrrrrrrr}
\tablecolumns{10}
\tablewidth{0pt}
\tablenum{6}
\tablecaption{Comparison of Oxygen Abundances}
\tablehead{
     \colhead{}&
     \colhead{}&
     \multicolumn{2}{c}{KPNO-02}&
     \colhead{}&
     \multicolumn{2}{c}{KPNO-04}&
     \colhead{}&
     \multicolumn{2}{c}{McD-04}\\
     \cline{3-4} \cline{6-7} \cline{9-10}\\
     \colhead{HIP}&
     \colhead{}&
     \colhead{[O/H]}&
     \colhead{$\sigma$}&
     \colhead{}&
     \colhead{[O/H]}&
     \colhead{$\sigma$}&
     \colhead{}&
     \colhead{[O/H]}&
     \colhead{$\sigma$}
     }
\startdata
14976 && 0.24  &$\pm 0.07$&&\nodata&\nodata   && 0.27  &$\pm 0.06$\\
16908 &&\nodata&\nodata   && 0.53  &$\pm 0.12$&& 0.40  &$\pm 0.12$\\
19148 && 0.16  &$\pm 0.05$&& 0.22  &$\pm 0.06$&&\nodata&\nodata   \\
19793 && 0.33  &$\pm 0.05$&&\nodata&\nodata   && 0.31  &$\pm 0.06$\\
19796 && 0.40  &$\pm 0.08$&&\nodata&\nodata   && 0.32  &$\pm 0.05$\\
20146 && 0.30  &$\pm 0.06$&&\nodata&\nodata   && 0.26  &$\pm 0.06$\\
20712 &&\nodata&\nodata   && 0.45  &$\pm 0.07$&& 0.37  &$\pm 0.05$\\
21099 && 0.15  &$\pm 0.07$&&\nodata&\nodata   && 0.22  &$\pm 0.06$\\
21112 && 0.24  &$\pm 0.05$&& 0.25  &$\pm 0.05$&&\nodata&\nodata   \\
\enddata
\end{deluxetable}

\begin{deluxetable}{lrrrrrrrrrrrrrr}
\tablecolumns{15}
\tablewidth{0pt}
\tablenum{7}
\tablecaption{Spot Synthesis Results}
\tablehead{
     \colhead{}&
     \colhead{}&
     \colhead{}&
     \colhead{}&
     \colhead{}&
     \colhead{}&
     \colhead{}&
     \colhead{}&
     \multicolumn{3}{c}{Observed}&
     \colhead{}&
     \multicolumn{3}{c}{Composite}\\
     \cline{9-11}  \cline{13-15}\\
     \colhead{}&
     \colhead{}&
     \colhead{$T_{\mathrm{eff}}$}&
     \colhead{}&
     \colhead{$T_{\mathrm{cool}}$}&
     \colhead{$T_{\mathrm{hot}}$}&
     \colhead{$T_{\mathrm{star}}$}&
     \colhead{}&
     \colhead{$\mathrm{EW}_{7771}$}&
     \colhead{$\mathrm{EW}_{7774}$}&
     \colhead{$\mathrm{EW}_{7775}$}&
     \colhead{}&
     \colhead{$\mathrm{EW}_{7771}$}&
     \colhead{$\mathrm{EW}_{7774}$}&
     \colhead{$\mathrm{EW}_{7775}$}\\
     \colhead{HIP}&
     \colhead{}&
     \colhead{(K)}&
     \colhead{}&
     \colhead{(K)}&
     \colhead{(K)}&
     \colhead{(K)}&
     \colhead{}&
     \colhead{(m{\AA})}&
     \colhead{(m{\AA})}&
     \colhead{(m{\AA})}&
     \colhead{}&
     \colhead{(m{\AA})}&
     \colhead{(m{\AA})}&
     \colhead{(m{\AA})}
     }
\startdata
18327 && 4988 && 4200 & 5900 & 4785 && 36.3  & 33.5 & 25.9 && 39.9  & 32.3 & 27.1\\
20082 && 4784 && 4000 & 5700 & 4569 &&\nodata& 25.8 & 18.0 &&\nodata& 25.1 & 21.4\\
18322 && 4573 && 3800 & 5500 & 4339 &&\nodata& 18.6 & 15.9 &&\nodata& 19.6 & 17.3\\
\enddata
\end{deluxetable}

\end{document}